\documentclass[aip,jmp,amsmath,amssymb,preprint]{revtex4-1}

\usepackage{graphicx}
\usepackage{dcolumn}
\usepackage{bm}

\usepackage{amsmath}
\usepackage{amssymb}
\usepackage{amsfonts}
\usepackage{mathrsfs}
\usepackage{epsfig}
\usepackage{bm}

\begin{document}

\title{Ion motion in the wake driven by long particle bunches in plasmas}

\author{J. Vieira$^1$, R.A. Fonseca$^{1,2}$, W.B. Mori$^3$, L.O. Silva}
\affiliation{$^1$ GoLP/Instituto de Plasmas e Fus\~ao Nuclear, Instituto Superior Técnico, Universidade de Lisboa, Portugal\\
$^2$ DCTI, ISCTE - Instituto Universit\'ario de Lisboa, Portugal\\
$^3$ Department of Physics and Astronomy, University of California, Los Angeles~CA, USA.}

\begin{abstract}
We explore the role of the background plasma ion motion in self-modulated plasma wakefield accelerators. We employ J. Dawson's plasma sheet model to derive expressions for the transverse plasma electric field and ponderomotive force in the narrow bunch limit. We use these results to determine the on-set of the ion dynamics, and demonstrate that the ion motion could occur in self-modulated plasma wakefield accelerators. Simulations show the motion of the plasma ions can lead to the early suppression of the self-modulation instability and of the accelerating fields. The background plasma ion motion can nevertheless be fully mitigated by using plasmas with heavier plasmas.
\end{abstract}


\maketitle

\section{\label{sec:intro}Introduction}

Particle acceleration plays an important role in several applications, ranging from science and medicine to industry. These applications require accelerators capable to deliver high energy electrons and ions in a controllable and reproducible fashion. In order to reach the required stability, conventional accelerators operate with accelerating gradients below material breakdown thresholds, which are on the order of a few tens-hundreds of MV/m. The low accelerating gradients imposed by material breakdown thresholds, however, result in very long acceleration distances. For instance the future International Linear Collider (ILC) will require more than 30 Km to accelerate 500 GeV electron/positron bunches. Because the size of the accelerator also determines its cost, there is a strong interest to develop advanced techniques that could drastically reduce the dimensions of particle accelerators. Plasmas are attractive to this purpose because they can sustain very large electric fields in excess of $\mathrm{E_{accel}~[V/m]} \simeq 0.96\sqrt{n_0\mathrm{[cm^{-3}]}}$ where $n_0$ is the plasma density. As a result, the potential of using plasmas to build compact ($1-100~\mathrm{m}$), high energy gain ($10-100~\mathrm{GeV}$) accelerators is currently being explored.

Plasma accelerators use particle bunches (Plasma Wakefield Accelerator - PWFA~\cite{bib:chen_prl_1985}) or laser pulses (Laser Wakefield Accelerator - LWFA~\cite{bib:tajima_prl_1979}) to drive relativistic plasma waves with large amplitude longitudinal electric fields that can be used to accelerate particles to high energies. First plasma acceleration experiments were performed in the limit where the driver was much longer than the plasma wavelength. Several configurations were then proposed in this limit and in the linear regime, including laser beatwave excitation~\cite{bib:beatwave}, wake excitation by the laser self-modulation instability~\cite{bib:lsmi}, and resonant wakefields driven by a train of electron bunches~\cite{bib:etrain} or laser pulses~\cite{bib:ltrain}. More recent experiments, however, use high energy drivers (1-100 J) much shorter than the plasma wavelength (i.e. with durations ranging between $30-50~\mathrm{fs}$). These particle bunch or laser pulse drivers can expel nearly all plasma electrons at its passage, exciting strongly non-linear wakefields in the bubble or blowout regime~\cite{bib:lu_2006}. Experiments in the blowout regime regime demonstrated self-trapping and acceleration of plasma electrons to a few GeV in less than 10 cm in the LWFA~\cite{bib:lwfa}, and acceleration of a fraction of driving bunch electrons to 40 GeV in less than a meter in the PWFA~\cite{bib:pwfa}.

There are several challenges that need to be addressed to advance this technology towards applications. One of them is to use higher energy drivers to enhance electron energy gains to 10-100 GeVs. Another is to bring repetition rates and overall stability closer to conventional accelerators. The high energy, high stability, and high repetition rates of the proton bunches provided by the Large-Hadron-Collider at CERN could meet these challenges and are therefore attractive as drivers for plasma acceleration. Numerical simulations showed that compressed, $200~\mu m$ long LHC-like proton bunches could excite wakefields in the blowout regime, leading to the acceleration of 500 GeV electron bunches in 500 meter long $n_0\simeq 10^{14}-10^{16}~\mathrm{cm}^{-3}$ plasmas in a single stage. This is called the proton-driven plasma wakefield accelerator - PDPWFA~\cite{bib:caldwell_natphys_2009}.

Available proton bunches, however, are much longer than the plasma wavelength ($\lambda_p$) with lengths on the order of $\sigma_z=10~\mathrm{cm}$. In addition, corresponding proton bunch charge densities ($n_b\simeq 10^{12}~\mathrm{cm}^{-3}$) are much lower than typical plasma densities for plasma acceleration. Therefore, the non-linear blowout regime will not be reached. However, large amplitude wakefields can still be produced through the self-modulation instability (SMI)~\cite{bib:kumar_prl_2010}. There are many analogies between self-modulation of particle bunches and laser pulses~\cite{bib:mori_ieee_1997} and these scenarios can also be relevant for the propagation of intense streams in astrophysics~\cite{bib:silva_aip_2006}. The self-modulation instability amplifies initial bunch density and bunch radius modulations at the plasma wavelength ($\lambda_p$). In turn, these modulations enhance the plasma wave amplitude, reinforcing the rate at which self-modulation occurs. When the bunch becomes fully self-modulated, it excites large amplitude plasma waves that grow from the head to the tail of the bunch. These wakefields can then be used to accelerate electrons or positrons. 

It has been shown that the wake phase velocity during SMI growth is much smaller than the driver velocity~\cite{bib:schroeder_prl_2011}. This can limit the energy gain by externally injected particle bunches. After SMI saturation, however, the wake phase velocity is identical to the driver velocity~\cite{bib:pukhov_prl_2011}. Ideal conditions for particle acceleration are then met after the saturation of SMI. The potential of self-modulated plasma accelerators to accelerate electrons to high energies after the saturation of the SMI has lead to the design of several experiments at CERN, SLAC~\cite{bib:vieira_pop_2012} and at other laboratories. Moreover, similar configurations using trains of laser pulses are also being considered~\cite{bib:ltrain}.

The length of the long bunches required in future self-modulation experiments can be comparable to the background plasma ion wavelength. If it occurs, the background plasma ion motion can lead to SMI suppression, strongly damping plasma wakefields, and inhibiting particle acceleration~\cite{ionmotion} even in the linear regime, where self-modulation experiments are likely to operate. It is thus important to avoid the motion of background plasma ions in experiments by, for instance, using plasmas with heavier ions~\cite{bib:vieira_prl_2012}. This was also suggested as a means to suppress ion dynamics in non-linear wakefields driven by short drivers~\cite{bib:rosensweig_prl_2005}. Interestingly, it has been shown that suitable control of the ion dynamics in the blowout regime can preserve the emittance of accelerated bunches~\cite{bib:muggli_ion}.

In this paper we examine the background plasma ion dynamics in conditions relevant for self-modulated wakefield accelerators. Our work can also be extended for wake excitation by trains of particle bunches. Our model is strictly valid in the narrow bunch limit, where future experiments are likely to operate, and where the plasma electrons trajectories are mostly determined by the radial plasma electric fields~\cite{bib:vieira_prl_2012}. Specifically, we demonstrate that linear wakefield excitation theory analytical expressions do not accurately reproduce particle-in-cell (PIC) simulations in the narrow driver limit. We then derive exact expressions for the transverse plasma electric field in this limit (i.e. where future proton self-modulated PDPWFA experiments will operate) using the plasma sheet model. These expressions, which accurately reproduce PIC simulation results, reveal that the wakefields can be strongly anharmonic even in the linear regime. The anharmonicities are due to the variations of the amplitude of the radial/transverse plasma electron oscillations across the driver. We then find generalized expressions for the plasma ponderomotive force and use these expressions to determine the ion density perturbations driven by the plasma ponderomotive force in the narrow driver limit. Our analytical findings are in agreement with PIC simulation results. Finally, we show that the motion of the background plasma ions in self-modulated plasma wakefield regimes can suppress self-modulation and particle acceleration. However, the ion motion can be mitigated by using plasmas with higher atomic numbers. 

This paper is organised as follows. In Sec.~\ref{sec:sec2} we determine the transverse/radial trajectories of plasma electron sheets in cylindrical and cartesian 2D-slab geometry using J. Dawson's plasma sheet model~\cite{bib:dawson_pr_1959}. Analytical results are confirmed by particle-in-cell (PIC) simulations in Osiris~\cite{bib:osiris}. In Sec.~\ref{sec:sec3} we derive generalised expressions for the plasma ponderomotive force that can act on the plasma ions. Theoretical results agree with simulations until the onset of fine-scale mixing of electron trajectories and wave breaking. In Sec.~\ref{sec:sec4} we determine the ion density perturbations driven by the plasma ponderomotive force and compare results with PIC simulations in conditions relevant for the PDPWFA. Finally in Sec.~\ref{sec:sec5} we present the conclusions.

\section{Transverse plasma waves in the narrow driver limit}
\label{sec:sec2}

We study wakefield excitation in the narrow bunch limit, where azimuthal magnetic fields can be neglected~\cite{bib:vieira_prl_2012}. According to Dawson's sheet model~\cite{bib:dawson_pr_1959}, the non-relativistic equation of motion for an electron ring pushed by an external particle bunch driver in the narrow bunch limit is:
\begin{equation}
\label{eq:ering}
c^2 \frac{\mathrm{d}^2 r_e}{\mathrm{d} \xi^2} = -\frac{\omega_p^2 r_e}{2} + \frac{\omega_p^2 r_{e0}^2}{r} - \frac{e E_r^b}{m_e},
\end{equation}
where $\omega_p$ is the electron plasma frequency, $c$ the speed of light, $m_e$ and $e$ the electron mass and charge, $\xi=z-c t$ is the distance to the head of the particle bunch driver, $r_e(\xi)$ is the electron radial displacement, $r_{e0}=r_e(0)$ is the initial radial position of the electron, $E_r^b$ is radial electric field of a particle bunch driver ($e$ is the electron charge). The corresponding force acting on plasma electrons is $e E_r^b$. The first term on the right-hand-side of Eq.~(\ref{eq:ering}) is the ion channel attractive electric field ($E_i=(\omega_p^2/e m_e) r/2$) and the second, the electron repulsive field ($E_e=(\omega_p^2/e m_e) r_0^2/r$).

The electric field of the plasma wave, given by $E_{w}=E_i+E_e$, can be determined by first solving Eq.~(\ref{eq:ering}) yielding $r_e(r_{e0})$, then by inverting the electron trajectories giving $r_{e0}(r)=r^{-1}(r_e)$, and finally inserting the inverted electron trajectories back into the expression for $E_w$. In order to solve Eq.~(\ref{eq:ering}), we perform a Taylor expansion for small bunch displacements up to order $\mathcal{O}\left([r-r_0]/r_0\right)^2 \ll 1$ yielding:
\begin{eqnarray}
\label{eq:ering-taylor}
\frac{\mathrm{d}^2 \Delta r_e}{\mathrm{d} \xi^2} = & - &\frac{e E_r^b}{m}  - \left[\omega_p^2+\frac{\mathrm{d}}{\mathrm{d} r_{e0}}\left(\frac{e E_r^b}{m}\right) \right]\Delta r_e \nonumber \\ 
& - &\frac{1}{2}\left[\frac{\omega_p^2}{r_{e0}}+\frac{\mathrm{d}^2}{\mathrm{d} r_{e0}^2}\left(\frac{e E_r^b}{m}\right) \right]\Delta r_e^2,
\end{eqnarray}
where $\Delta r_e = r_e - r_{e0}$ is the displacement of the electron to its initial radial position. The $\mathcal{O}(\Delta r_e)^2$ term is absent in 2D slab geometries, but can be important in 3D scenarios. 

We first consider forced electron oscillations, neglecting the terms of order $\mathcal{O}(\Delta r_e)^2$. This approximation still retains most important important physical mechanisms relevant for the background ion dynamics. In these conditions  Eq.~(\ref{eq:ering-taylor}) becomes:
\begin{eqnarray}
\label{eq:ering-taylor-forced}
\frac{\mathrm{d}^2 \Delta r_<}{\mathrm{d} \xi^2} = - \frac{e E_r^b}{m} - \left[\omega_p^2+\frac{\mathrm{d}}{\mathrm{d} r_0}\left(\frac{e E_r^b}{m}\right) \right]\Delta r,
\end{eqnarray}
where $\Delta r_<=r_{e}-r_{e0}$ in the region of the driver. Equation~(\ref{eq:ering-taylor-forced}), also valid for 2D slab geometry, is an harmonic oscillator forced by the electrostatic fields of a charged particle bunch. According to Eq.~(\ref{eq:ering-taylor-forced}), electrons oscillate sinusoidally about an equilibrium position according to:
\begin{equation}
\label{eq:ering-forced-solution}
\Delta r_{<} = A_{<}(r_{e0}) \left[1-\cos\left(\phi_{<}\right)\right],
\end{equation}
where:
\begin{equation}
\label{eq:amplitude-forced}
A_{<}(r_{e0})=-\frac{E_r^b(r_{e0})}{m_e \omega_p^2/e+\nabla_{r_{e0}}\left[E_r^b(r_{e0})\right]},
\end{equation}
is the amplitude of the oscillation, and where:
\begin{equation}
\label{eq:phase-forced}
\phi_{<}(r_{e0})=\frac{\omega_p \xi}{c}\left[1+\frac{e}{\omega_p^2 m_e} \nabla_{r_{e0}}E_r^b(r_{e0})\right]^{1/2}, 
\end{equation}
is the phase of the oscillation. The phase and frequency of the oscillation depend on $r_{e0}$. Fine-scale mixing of the electron trajectories will then occur when the oscillations of adjacent plasma sheets become roughly $\pi/2$ out of phase~\cite{bib:dawson_pr_1959}. We stress that although Eq.~(\ref{eq:ering-taylor}) can be used to predict when trajectory (sheet) crossing occurs, it is no longer valid afterwards. Equations~(\ref{eq:ering-forced-solution}) and (\ref{eq:phase-forced}) reveal that sheet crossing is due to the finite transverse gradients of the particle bunch. This contrasts with Ref.~\cite{bib:dawson_pr_1959}, where sheet crossing only occurs in cylindrical geometry because plasma oscillations are anharmonic. 

In order to derive expressions for free electron oscillations we set $E_r^b=0$ in Eq.~(\ref{eq:ering-taylor}) yielding:
\begin{equation}
\label{eq:ering-taylor-free}
\frac{\mathrm{d}^2 \Delta r_>}{\mathrm{d} \xi^2} = - \omega_p^2 \Delta r_> + \frac{\omega_p^2 \Delta r_>^2}{2 r_{e0}},
\end{equation}
where $\Delta r_>=r_e-r_{e0}$ in the regions absent of driver. By keeping the term $\mathcal{O}(\Delta r_>^2)$ we can understand the role of the an-harmonicities in the structure of the wakefield and in the motion of the background plasma ions. Note that Eq.~(\ref{eq:ering-taylor-free}) is only valid for 2D cylindrical geometry because the term $\mathcal{O}(\Delta r_>)^2$ is absent from 2D slab geometries, where free plasma oscillations are purely harmonic.

We can solve Eq.~(\ref{eq:ering-taylor}) using a pertubative technique by looking for solutions $\Delta r_>=\sum_{i=0}^j \epsilon^i \left(\Delta r_>\right)_i$, where $\epsilon_i$ is a small parameter. Collecting terms with similar order in $\epsilon$ for the first two terms of the latter expansion ($j=2$) yields the following two coupled differential equations for $(\Delta r_>)_1$ and $(\Delta r_>)_2$:
\begin{eqnarray}
\label{eq:ering-coupled}
\frac{\mathrm{d}^2 (\Delta r_>)_1}{\mathrm{d} \xi^2} & = & - \omega_p^2 (\Delta r_>)_1 \\
\frac{\mathrm{d}^2 (\Delta r_>)_2}{\mathrm{d} \xi^2} & = & -\omega_p^2 \left((\Delta r_>)_2-\frac{(\Delta r_>)_1^2}{2 r_0}\right)
\end{eqnarray}
with solution:
\begin{equation}
\label{eq:ering-solution}
\Delta r_> = A_>(r_{e0}) \cos\left( \phi_> \right) -\frac{A_>^2(r_{e0})}{12 r_0}\left[\cos\left(2\phi_>\right)-3\right]+\mathcal{O}\left(A_>^3\right),
\end{equation}
where $A_>(r_{e0})$ is the amplitude of the electron ring oscillation, $\Delta r_>=r_e-r_{e0}$, and where 
\begin{equation}
\label{eq:phase-free}
\phi_> = \omega_p \xi\left(1+\frac{1}{12}\frac{A_>(r_{e0})^2}{r_{e0}^2}\right),
\end{equation}
is its phase including the lowest order non-linear correction to the plasma frequency. If plasma electrons are driven by a flat-top driver with length $\sigma_z$ then after the driver has passed:
\begin{equation}
\label{eq:amplitude-free}
A_{>}(r_{e0})=\left[\Delta r_>(\sigma_z)^2+\frac{c^2}{\omega_p^2} \left(\frac{\mathrm{d} \Delta r_>(\sigma_z)}{\mathrm{d} \xi}\right)^2\right]^{1/2},
\end{equation}
where $\Delta r_>(\sigma_z) = r_{e}(\sigma_z)-r_{e0}$. Equation~(\ref{eq:phase-free}) shows that the electron oscillations are anharmonic because their frequency depends on their amplitude. Hence, fine scale-mixing of the electron trajectories occurs as in Ref.~\cite{bib:dawson_pr_1959}.

We compared these predictions with 2D PIC simulations in Osiris for the forced plasma oscillations scenario given by Eq.~(\ref{eq:ering-forced-solution}). Simulations use an infinite and external non-evolving transverse electric field to excite the plasma, given by:
\begin{equation}
\label{eq:driver}
E_r^b = \frac{E_{0} m_e \omega_p^2}{e} x_{\perp} \exp\left(-\frac{x_{\perp}^2}{\sigma_{\perp}^2}\right),
\end{equation}
where $x_{\perp}$ is the transverse position and where $E_0 = 0.3 $, such that the amplitude of the plasma electron trajectories respects $\Delta r_</r_{e0}\ll 1$. In addition, $\sigma_r = 0.1 c/\omega_p\ll c/\omega_p$, ensuring that wakefield excitation occurs in the narrow bunch limit. Simulations use a computation box traveling at $c$ with dimensions $100\times1.26~\left(\mathrm{c/\omega_p}\right)^2$ in the longitudinal and transverse directions divided into $5120\times600$ cells with $4\times4$ particles-per-cell.

\begin{figure}
\centering\includegraphics[width=0.7 \columnwidth]{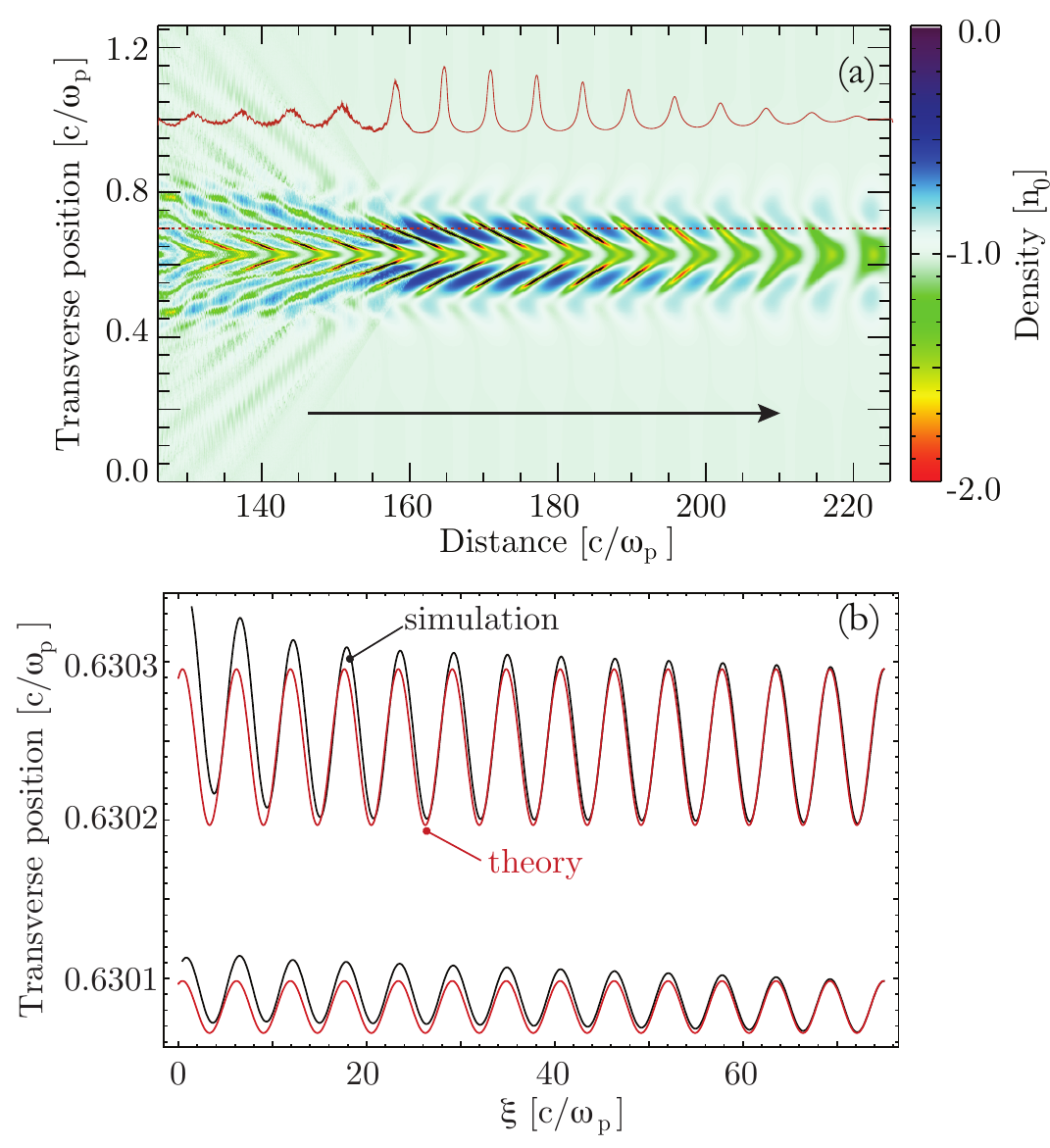}
\caption{Simulation results illustrating the plasma response to a long external electric field driver to mimic the presence of a particle bunch. The external electric field is given by $E_r^b=(m_e \omega_p^2 E_0/e) x_{\perp} \exp\left[-x_{\perp}^2/\sigma_{\perp}^2\right]$ with $E_0=0.3$ and $\sigma_{\perp}=0.1 c/\omega_p$. (a) shows the resulting plasma electron density. The solid red-line is a line-out taken at the dashed red line highlighting the strong plasma wave non-linearity even in linear regimes. (b) Comparison between 2D simulation and theoretical results for plasma electron trajectories. Plasma oscillations are excited by an external electric field driver to mimic the presence of a particle beam on the form . Simulation results are in black and theoretical results in red. The direction of the arrow indicates the direction of propagation}
\label{fig:plasma}
\end{figure}

Simulation results are illustrated in Fig.~\ref{fig:plasma}a.The plasma is initially outside the simulation window. Electrons then start oscillating as soon as the moving window reaches the plasma. In this setup, the plasma electron density wavelength corresponds to the distance traveled by the simulation box during approximately a plasma period. Since the simulation box travels at $c$, this wavelength is close to the plasma wavelength. We also note that Fig.~\ref{fig:plasma}a shows the plasma response for a fixed $t$. Thus $\xi$ plays the role of $z$ at a fixed time. Figure~\ref{fig:plasma}, which illustrates the excitation of non-linear plasma waves, shows that plasma density wavefronts bend backwards away from the axis. In agreement with Eq.~(\ref{eq:phase-forced}), this indicates that the plasma wavelength decreases when going away from the axis. The plasma wavelength variations across the driver lead to strongly non-linear plasma density perturbations, even though electron displacements are much smaller than the plasma skin-depth. For instance the lineout on Fig.~\ref{fig:plasma}a illustrates the formation of electron density buckets and density spikes off-axis. Electron trajectories taken at different transverse positions, including regions where wakefields are non-linear, are 
shown in Fig.~\ref{fig:plasma}b. Simulation results are in very good agreement with theoretical predictions given by Eq.~(\ref{eq:ering-taylor-forced}), except at the end of the simulation box, where theory underestimates electron amplitudes of oscillations by $\approx 30\%$. Simulations suggest that this is due to the fine-scale mixing of the plasma electron trajectories, where Eq.~(\ref{eq:ering}) is not valid. Before wave breaking occurs, Fig.~\ref{fig:plasma}b then shows that electron trajectories are purely sinusoidal, and indicates that the non-linear wakefield structures are due to both $A_<$ and $\phi_<$ being functions of $r_{e0}$, i.e. that the amplitude and frequency of electron oscillations varies across the driver. 

After studying the overall plasma dynamics and plasma electron trajectories, we now determine the radial plasma electric field by inverting the expressions for the electron trajectory inside the driver, $\Delta r_<$, given by Eq.~(\ref{eq:ering-forced-solution}), and after the driver has passed, $\Delta r_>$, given by Eq.~(\ref{eq:ering-solution}). The inversion of plasma electron trajectories is performed assuming small radial displacements, in agreement with the conditions leading to Eq.~(\ref{eq:ering-taylor}). We then expand the electron trajectories $r_e=R_e(r_{e0})$ in Taylor series for small $\Delta r_e$ yielding to first order in $\mathcal{O}(r_{e0}-r_e)$:
\begin{equation}
\label{eq:expansion-electrons}
R_e(r_{e0}) = r_e = R_e(r_e) + \left(r_{e0}-r_e\right)\nabla_{r_{e0}=r_e} R_e.
\end{equation}
The inverted electron trajectories are thus given by:
\begin{equation}
\label{eq:inverted-expansion-electrons}
r_{e0} = r_e + \frac{r_e-R_e}{\nabla_{r_{e0}}R_e},
\end{equation}
where the gradient is evaluated at $r_{e0}=r_e$.

We start by determining the inverted trajectories for the forced oscillation scenario. We then insert Eq.~(\ref{eq:ering-forced-solution}) for $\Delta r_<$ into Eq.~(\ref{eq:inverted-expansion-electrons}) and neglect the gradients associated with the phase of the electron oscillations ($\mathrm{d}\phi_</\mathrm{d} r_{e0}=0$). Then, $r_{e0}$ becomes:
\begin{equation}
\label{eq:ering-forced-inverted}
r_{e0} = r_e - \frac{A_< \left[1-\cos\left(\phi_<\right)\right]}{1+\nabla_{r_e}A_<\left[1-\cos\left(\phi_<\right)\right]},
\end{equation}
where all derivatives are evaluated at $r_{e0}=r_e$. Similarly, the inverted trajectories for the free oscillation scenario are given by inserting Eq.~(\ref{eq:ering-solution}) for $\Delta r_>$ into Eq.~(\ref{eq:inverted-expansion-electrons}) yielding:
\begin{equation}
\label{eq:ering-free-inverted}
r_{e0} = r_e + \frac{-A_> \cos\left(\phi_>\right) + \frac{A_>^2}{12 r_e} \left[\cos\left(2 \phi_>\right)-3\right]}{1+\cos\left(\phi_>\right)\nabla_{r_e} A_> + \frac{1}{12}\left[\cos\left(2 \phi_>\right)-3\right]\left(\frac{A_>^2}{r_e^2}-\frac{\nabla_{r_e} A_>^2}{r_e}\right)},
\end{equation}
where the derivatives are evaluated at $r_{e0}=r_e$. Note that Eqs~(\ref{eq:ering-forced-inverted}) and (\ref{eq:ering-free-inverted}) can also be derived using Lagrange implicit function theorem.

Using Eqs.~(\ref{eq:ering-forced-inverted}) and (\ref{eq:ering-free-inverted}) we can now find the transverse plasma electric field by replacing $r_{e0}$ into $E_{r}=E_i+E_e+E_r^b$. For the forced oscillation case this yields:
\begin{equation}
\label{eq:efield-plasma-forced}
E_< = \frac{\hat{E}_<\left(1-\cos{\phi_<}\right)}{1+\frac{e \nabla_r \hat{E}_<}{m_e\omega_p^2}\left(1-\cos{\phi_<}\right)}+E_r^b,
\end{equation}
where $\hat{E}_<=m_e\omega_p^2 A_</e$ is the amplitude of the transverse electric field of the plasma wave. For wide wakefields such that $e/(m_e \omega_p^2)\nabla_r \hat{E}_< \ll 1$, Eq.(\ref{eq:efield-plasma-forced}) recovers linear wakefield theory analytical results~\cite{bib:lu_pop_2005}. For narrow wakes such that $e/(m_e \omega_p^2)\nabla_r \hat{E}_< \gtrsim 1$, Eq.~(\ref{eq:efield-plasma-forced}) provides anharmonic corrections to the amplitude of the wakefield.

The expression for $E_>$ in the case of free plasma oscillations including anharmonic corrections associated with the 2D cylindrical geometry is cumbersome. Simpler formulas can, however,  be found by expanding $E_>$ into $\mathcal{O}\left(\hat{E}_>\right)^2$ where $\hat{E}_>=m_e\omega_p^2 A_>/e$ yielding:
\begin{eqnarray}
\label{eq:efield}
E_> & = & \frac{\hat{E}_>\cos\left(\phi_>\right)}{1+\cos\left(\phi_>\right) e \nabla_r \hat{E}_>/(m_e \omega_p^2)} \nonumber \\
& - & \left(\frac{ 6 \cos\left(\phi_>\right)^2+\left(-3+\cos\left(2\phi_>\right)\right)\left(1-\cos\left(\phi_>\right) e \nabla_r \hat{E}_>/(m_e \omega_p^2)\right)}{12 r \left[1+\cos\left(\phi_>\right) e \nabla_r \hat{E}_>/(m_e \omega_p^2)\right]^2}\right)\frac{e \hat{E}_>^2}{m_e c \omega_p}.
\end{eqnarray}  
Linear wakefield excitation theory is recovered from Eq.~(\ref{eq:efield}) in the limit of wide plasma waves compared with the plasma skin depth such that $(e/m_e\omega_p^2) \nabla_r \hat{E}_>\ll 1$ and also for $m_e c \omega_p \hat{E}_> / e\ll 1$. The second term of Eq.~(\ref{eq:efield}) is a correction associated with anharmonic plasma electrons trajectories in cylindrical geometry. We stress that Eq.~(\ref{eq:efield}) shows that the transverse plasma electric field can become strongly non-linear when $(e/m_e\omega_p^2) \nabla_r \hat{E}_>\gtrsim 1$ even for small transverse (radial) electron displacements, $\Delta r_e/r_{e0}\ll 1$.

\begin{figure}
\centering\includegraphics[width=\columnwidth]{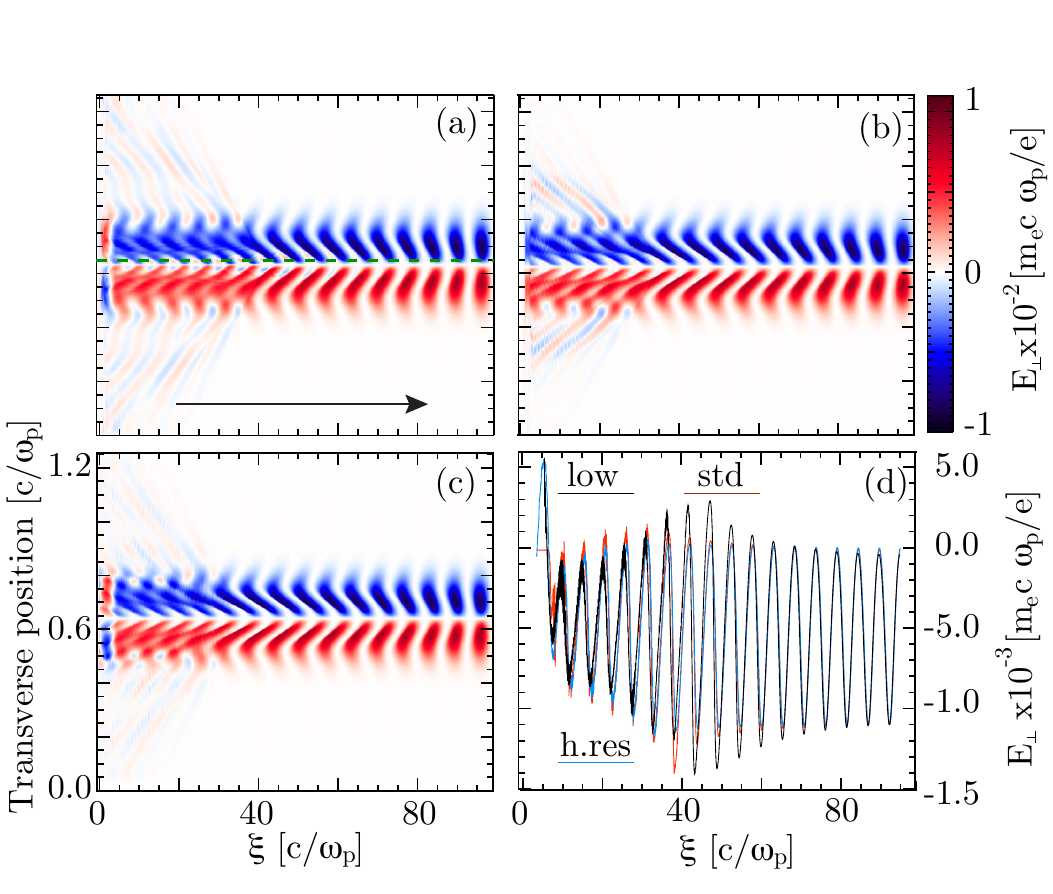}
\caption{Simulation results showing the transverse electric field profile. The plasma is excited by an external electric field given by $e E_r^b/(m_e c\omega_p) = 0.2 \exp\left(-x_{\perp}^2/\sigma_{\perp} \right)$ with $\sigma_{\perp}=0.1c/\omega_p$. The driver moves from left to right as indicated by the arrow. (a), (b), and (c) show results from simulations using 30, 60 and 120 points per $\sigma_{\perp}$ in the transverse direction, respectively, and $330$ points per $\lambda_p$ in the longitudinal direction. (d) shows a comparison between lineouts of the field took at $x_{\perp}=0.5\sigma_{\perp}$, where low, std, and high refers to simulations using 30, 60 and 120 points per $\sigma_{\perp}$, respectively.}
\label{fig:convergence}
\end{figure}

In order to compare theoretical results given by Eq.~(\ref{eq:efield-plasma-forced}) with numerical simulations, we first performed a numerical resolution parameter scan. Figures~\ref{fig:convergence}a-c then show simulation results with 320 points per plasma wavelength in the longitudinal direction and 30 (Fig.~\ref{fig:convergence}a), 60 (Fig.~\ref{fig:convergence}b), 120 (Fig.~\ref{fig:convergence}c) points per driver transverse spot-size $\sigma_{\perp}$ in the transverse directions. More important differences occur for the resolutions of Figs.~\ref{fig:convergence}a-b near the axis and for $\xi \lesssim 40-60~c/\omega_p$, while Figs.~\ref{fig:convergence}b-c are nearly identical. These results show that accurate modelling of the wakefield structures near the axis (i.e. in regions where the background plasma ion motion would be stronger) require at least 60 points per $\sigma_{\perp}$. These simulations also showed that the use of lower resolutions results in higher wakefield amplitudes (Fig.~\ref{fig:convergence}d).

\begin{figure}
\centering\includegraphics[width=0.8 \columnwidth]{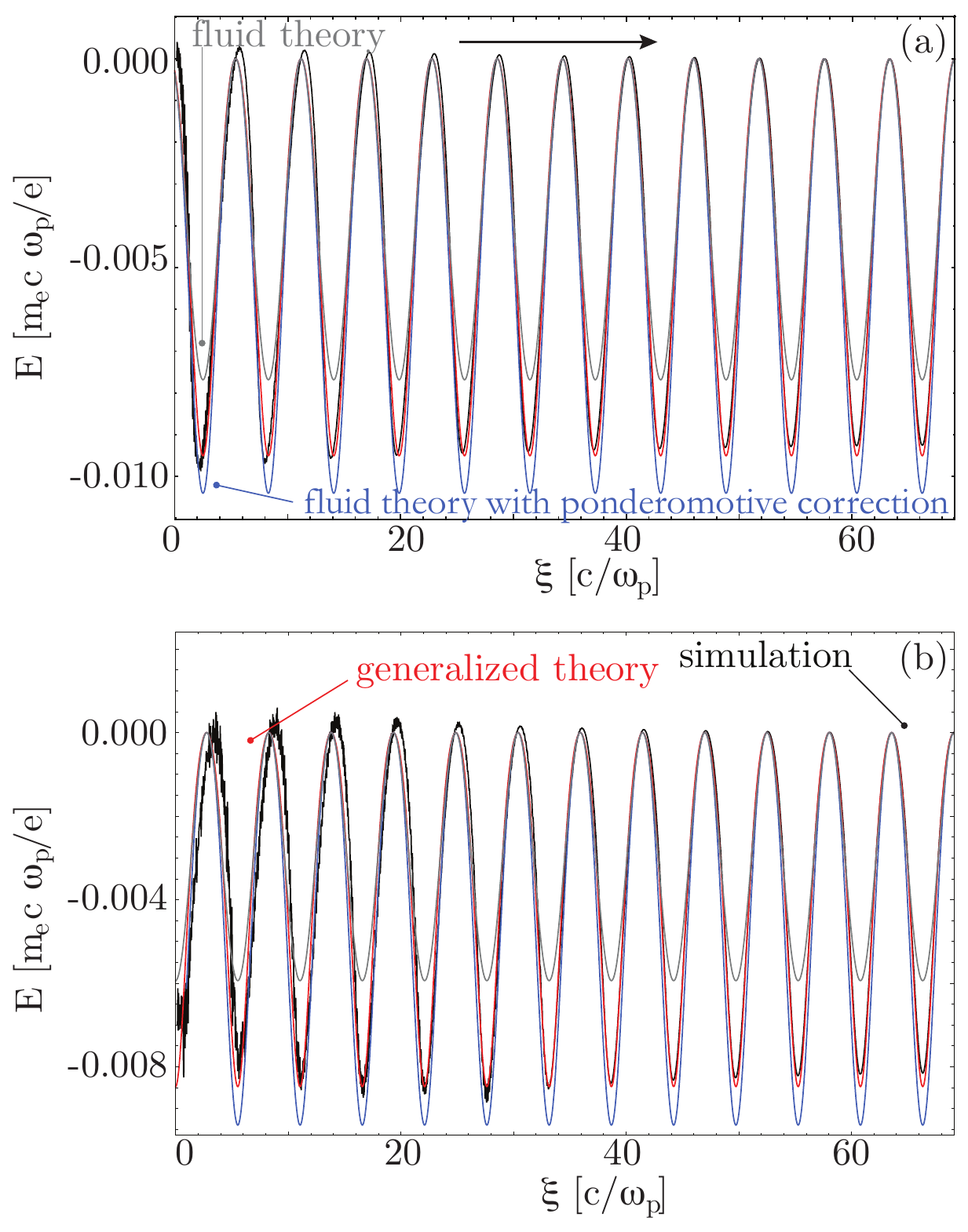}
\caption{Comparison between 2D simulation and theory for the transverse plasma electric field. The plasma oscillations are excited by an external electric field driver to mimic the presence of a particle beam on the form $e E_r^b/(m_e c\omega_p)=E_0 x_{\perp} \exp\left[-x_{\perp}^2/\sigma_{\perp}^2\right]$. The driver moves from left to right as indicated by the arrow. (a) uses $E_0=0.2$ and the fields are took at $x_{\perp}=0.1\sigma_{\perp}$ and (b) $E_0=0.3$ and fields are took at $x_{\perp}=0.2\sigma_{\perp}$. Simulation results are in black and the generalised theory developed in the paper is in red. Linear theory results are shown by the dashed-gray lines, and linear theory results including a first order correction associated with the plasma ponderomotive force are shown in solid-gray.}
\label{fig:field-comp}
\end{figure}

We compare theory and simulations for the transverse wakefield structure in Fig.~\ref{fig:field-comp}. The simulation used 120 points per $\sigma_{\perp}$, with $E_0=0.2$ (Fig.~\ref{fig:field-comp}a) and $E_0=0.3$ (Figure~\ref{fig:field-comp}b). Figure~\ref{fig:field-comp} also shows the predictions of linear wakefield excitation theory including a first order correction due to the plasma ponderomotive force, and the predictions of Eq.~(\ref{eq:efield-plasma-forced}). We found very good agreement between simulation results and Eq.~(\ref{eq:efield-plasma-forced}) before wave-breaking occurs, i.e. for $\xi\lesssim 20~c/\omega_p$. We also found that linear theory results underestimate the amplitude of the simulations by about $20\%$. However, when including corrections associated with the plasma ponderomotive force, linear theory then overestimates simulation results by less than $10\%$. Unlike in 1D scenarios, simulations shown in Figs.~\ref{fig:plasma},\ref{fig:convergence} and \ref{fig:field-comp} then demonstrate that trajectory crossing leading to wave breaking always occurs in multi-dimensions, and that it can occur even when the plasma electron trajectories is purely sinusoidal with small amplitude of oscillation such that $\Delta r_e/r_{e0}\ll 1$.

\section{Generalised plasma wave transverse ponderomotive force}
\label{sec:sec3}

We can now determine the transverse (i.e. ponderomotive force) force felt by background plasma ions due to the plasma electron transverse wakefield. For narrow and long drivers, the transverse components of the ponderomotive force are dominant because transverse gradients are much larger than longitudinal gradients. Since the longitudinal electric field is also much smaller than the radial electric field in the narrow driver limit, the motion of the background plasma ions is preferably in the radial direction. Thus, to find the force acting on the plasma ions we now determine $E_r$ averaged on a plasma period using Eq.~(\ref{eq:efield-plasma-forced}), for the forced electron oscillations inside the driver, and Eq. (\ref{eq:efield}), for free electron oscillations after the driver has passed. The average transverse electric field inside the driver is: 
\begin{eqnarray}
\label{eq:efield-forced-average}
\langle E_< \rangle & = & \frac{1}{2 \pi}\int_0^{2\pi}E_< \mathrm{d}\phi = \frac{\hat{E}_<}{e \nabla_r \hat{E}_</(m_e\omega_p^2)} \nonumber \\
& \times &\left(1-\frac{1}{\sqrt{1-2 e\nabla_r \hat{E}_</(m_e \omega_p^2)}}\right) + E_r^b. 
\end{eqnarray}
Equation~(\ref{eq:efield-forced-average}) generalises linear theory predictions for the transverse plasma wave ponderomotive force to the narrow bunch limit, and differs from linear wakefield theory in the narrow bunch limit, where $(m_e \omega_p^2/e)\nabla_r \hat{E}_< \gtrsim 1$. Linear theory results can then be recovered in the limit of wide driver bunches, when $(m_e \omega_p^2/e)\nabla_r \hat{E}_< \ll 1$. Direct connection with linear theory in the limit of wide drivers can be established by recalling the definition for $E_<$ and $A_<$, and by expanding Eq.~(\ref{eq:efield-forced-average}) into first order in powers of $\mathcal{O}\left(\hat{E}_<\right)$ yielding:
\begin{equation}
\label{eq:efield-forced-average-taylor}
\langle E_< \rangle = -\frac{1}{2}\nabla_r \left(E_r^b\right)^2. 
\end{equation}

The average transverse electric field after the driver has passed is:
\begin{eqnarray}
\label{eq:efield-average}
\langle E_> \rangle & = & \frac{1}{2 \pi}\int_0^{2\pi}E_>\mathrm{d}\phi = \frac{\hat{E}_>}{e \nabla_r \hat{E}_>/(m_e \omega_p^2)} \nonumber \\
& \times &\left(1-\frac{1}{\sqrt{1-\left(e \nabla_r \hat{E}_>/(m_e \omega_p^2)\right)^2}}\right),
\end{eqnarray}
where only the first term in the expression for $E_>$ given by Eq.~(\ref{eq:efield}) was considered. Equation~(\ref{eq:efield-average}) is also a generalisation of linear theory in the narrow bunch limit. Connection with linear theory can be thus obtained by expanding Eq.~(\ref{eq:efield-average}) to second order in powers of $\mathcal{O}[e\hat{E}_</(m_e \omega_p c)]$, $\mathcal{O}\left[\nabla_r e \hat{E}_>/(m_e\omega_p^2)\right]$ yielding:
\begin{equation}
\label{eq:pf-fluid}
\langle E_> \rangle = -\frac{1}{4}\nabla_r \frac{e\hat{E}_<^2}{m_e \omega_p^2} - \frac{3}{8}\frac{e^3}{m_e^3 \omega_p^6} \hat{E}_> \left(\nabla_r \hat{E}_>\right)^3 + \mathcal{O}\left(\nabla_r \hat{E}_>\right)^5, 
\end{equation}
where the first term corresponds to the plasma ponderomotive force derived from linear theory. The second term is a correction, which becomes important for narrow plasma waves.

\section{Onset of background plasma ion motion}
\label{sec:sec4}

To determine the background plasma ion density perturbations in 2D cylindrical and 2D cartesian/slab geometries we consider the transformation of the volume elements mapping the initial ion position ($r_{i0}$) to its current position ($r_i(\xi)$). The volume element in cylindrical coordinates in the $r_{i0}$ space is $\mathrm{d}V_{i0}=n_{i0}r_{i0}\mathrm{d}r_{i0}$ where $n_{i0}=n_0$ is the background plasma ion density. The volume element $\mathrm{d}V_{i0}$ is related to the volume element in the $r_i$ space through the Jacobian of the transformation mapping $r_{i0}$ into $r_i$~\cite{bib:dawson_pr_1959}. In 2D cylindrical coordinates, $\mathrm{d}V_{i0}=n_{i0} r_{i0} (\mathrm{d} r_{i0}/\mathrm{d}r_{i})  \mathrm{d} r_i  = n_i r_i \mathrm{d} r_i$ considering $\mathrm{d}r_{i0}=(\mathrm{d} r_{i0}/\mathrm{d} r_{i}) \mathrm{d} r_i$. In cylindrical coordinates the ion density is then given by:
\begin{equation}
\label{eq:ni-general-2dcyl}
n_i^{\mathrm{cyl}} = n_{i0}\frac{r_{i0}(r_i)}{r_i}\left(\frac{\mathrm{d}r_{i0}(r_i)}{\mathrm{d} r_{i}}\right),
\end{equation}
In slab geometry the volume element is $\mathrm{d}V_{i0}=n_{i0}\mathrm{d}r_{i0}=n_{i0} (\mathrm{d} r_{i0}/\mathrm{d} r_{i}) \mathrm{d} r_i  = n_i \mathrm{d} r_i$. Hence the ion density in 2D slab geometry is:
\begin{equation}
\label{eq:ni-general-slab}
n_i^{\mathrm{slab}} = n_{i0}\left(\frac{\mathrm{d}r_{i0}}{\mathrm{d} r_{i}}\right).
\end{equation}

In order to evaluate Eqs.~(\ref{eq:ni-general-2dcyl}) and (\ref{eq:ni-general-slab}), we first determine expressions for transverse (radial) trajectories of the background plasma ions, $r_i(r_{i0})$. The background plasma ions respond to the plasma ponderomotive force, $\langle E_< \rangle$ and $\langle E_> \rangle$, derived in Sec.~\ref{sec:sec3}, as long as the background plasma ion repulsive force is negligible and as long as the background plasma ion motion does not affect plasma electron oscillations. These assumptions, which are valid for $(n_i-n_0)/n_0\ll 1$, lead to: 
\begin{equation}
\label{eq:ri-general}
\frac{\mathrm{d}^2 r_i}{\mathrm{d} \xi^2} = \frac{Z e}{m_i c^2} \langle E_{\lessgtr} \rangle, 
\end{equation}
where $Z$ is the ion charge normalised to the electron charge $e$, and where $E_{\lessgtr}=E_<$ for $\xi<\sigma_z$ (driven plasma oscillations), and $E_{\lessgtr}=E_>$ for $\xi>\sigma_z$ (free plasma oscillations). We can find approximate solutions to Eq.~(\ref{eq:ri-general}) for $\xi \ll \lambda_{pi}=2\pi/\omega_{pi}$, where $\omega_{pi}=\omega_p (m_e/m_i)^{1/2}$, by expanding $r_i=\Sigma_{n} (r_i)_n \xi^n$, where $(r_i)_n$ are constant factors. Substituting $r_i=(r_i)_0 + (r_i)_2 \xi^2+\mathcal{O}(\xi^3)$ in Eq.~(\ref{eq:ri-general}) then leads to:
\begin{equation}
\label{eq:ri-early-general}
r_{i} = (r_i)_0 + \frac{\xi^2}{2} \frac{Z e}{m_i c^2} \langle E_{\lessgtr}\rangle|_{r=(r_i)_0} + \mathcal{O}\left(\xi^3\right),  
\end{equation}
where we assumed that that ions were at rest at $\xi=0$, and where $(r_i)_0=r_{i0}$ is the initial radial position of an ion ring. Equation~(\ref{eq:ri-early-general}) gives the radial trajectory of background plasma ions due to the average force provided by the transverse plasma wakefield for $\xi\ll\lambda_{pi}$. Inversion of the background plasma ion trajectories, required to find $n_i^{\mathrm{cyl}}$ [Eq.~(\ref{eq:ni-general-2dcyl})] and $n_i^{\mathrm{slab}}$ [Eq.~(\ref{eq:ni-general-slab})], can be performed by following the procedure outlined in Sec.~\ref{sec:sec2}, resulting in: 
\begin{equation}
\label{eq:expansion}
R_i(r_{i0})=R_i + \left(r_{i0}-r_i\right)\nabla_{r_{i0}=r_i} R_i.  
\end{equation}
The inverted background plasma ion trajectories is then found by combining Eq.~(\ref{eq:ri-early-general}) with Eq.~(\ref{eq:expansion}):
\begin{equation}
\label{eq:ri0}
r_{i0} \approx r_i + \frac{r_i-R_i(r_i)}{\nabla_r R_i(r_i)} = r_i - \frac{Z e}{m_i c^2}\frac{\xi^2}{2} \langle E_{\lessgtr}\rangle + \mathcal{O}\left(\xi^3\right)
\end{equation}
where both $\nabla_r R_i(r_i)=\mathrm{d} R_i/\mathrm{d} r_{i0}$ and $\langle E_{\lessgtr}\rangle$ are evaluated at $r_{i0}=r_i$. Inserting Eq.~(\ref{eq:ri0}) in the expression for the background plasma ion density in 2D Cylindrical geometry [Eq.~(\ref{eq:ni-general-2dcyl})] yields:
\begin{eqnarray}
\label{eq:ni-general-2dcyl-early}
n_i^\mathrm{cyl} & = & n_{i0}\left[ 1 - \frac{Z e }{m_i c^2}\frac{\xi^2}{2}\left(\frac{\langle E_{\lessgtr} \rangle}{r_i}+\frac{\mathrm{d}\langle E_{\lessgtr}\rangle}{\mathrm{d}r_i}\right)\right] \nonumber \\
& = & n_{i0}\left( 1 - \frac{Z e}{m_i c^2}\frac{\xi^2}{2}\nabla\cdot\langle E_{\lessgtr}\rangle\right).
\end{eqnarray}
The background plasma ion density in 2D slab geometry is found by inserting the inverted ion trajectories [Eq.~(\ref{eq:ri0})] into the expression for $n_i^{\mathrm{slab}}$ [Eq.~(\ref{eq:ni-general-slab})]:
\begin{eqnarray}
\label{eq:ni-general-slab-early}
n_i^\mathrm{slab} & = & n_{i0}\left[ 1 - \frac{Z e}{m_i c^2}\frac{\xi^2}{2}\frac{\mathrm{d}\langle E_{\lessgtr}\rangle}{\mathrm{d}r_i}\right] \nonumber \\
& = & n_{i0}\left( 1 - \frac{Z e}{m_i c^2}\frac{\xi^2}{2}\nabla\cdot\langle E_{\lessgtr}\rangle\right).
\end{eqnarray}
The divergence operator is $\nabla \cdot = (1/r)(\mathrm{d} /\mathrm{d} r) r $ in cylindrical geometry (Eq.~(\ref{eq:ni-general-2dcyl-early})) and $\nabla \cdot = \mathrm{d} /\mathrm{d} r $ in 2D slab geometry (Eq.~(\ref{eq:ni-general-slab-early})). Equations~(\ref{eq:ni-general-2dcyl-early}) and (\ref{eq:ni-general-slab-early}), valid as long as the plasma electron trajectories $\Delta r_<$ and $\Delta r_>$ are given by Eqs.~(\ref{eq:ering-forced-solution}) and (\ref{eq:ering-solution}), are general expressions for the early evolution of the background plasma ion density. Note also that Eqs.~(\ref{eq:ni-general-2dcyl-early}) and (\ref{eq:ni-general-slab-early}) are consistent with results from fluid theory~\cite{bib:vieira_prl_2012}. 

Explicit expressions for the background ion density can be obtained near the axis ($r \ll \sigma_r$) within the particle bunch for $\xi<\sigma_z$:
\begin{equation}
\label{eq:ni-cyl-onaxis}
n_i^{\xi<\sigma_z} = n_0\left\{1-\alpha_{\mathrm{2D/3D}} \frac{m_e Z \omega_p^2\xi^2}{m_i c^2}\left[1-\frac{1}{\sqrt{1-\frac{2 e}{m_e \omega_p^2}\nabla_r \hat{E}_<}}+\frac{e}{m_e \omega_p^2}\nabla_r E_r^b\right]\right\},
\end{equation}
and outside the particle bunch for $\xi>\sigma_z$:
\begin{equation}
\label{eq:ni-slab-onaxis}
n_i^{\xi>\sigma_z} = n_0\left[1- \alpha_{\mathrm{2D/3D}} \frac{m_e Z \omega_p^2 \xi^2}{m_i c^2}\left(1-\frac{1}{\sqrt{1-e^2 (\nabla_r \hat{E}_>)^2/(m_e^2 c^4)}}\right)\right],
\end{equation}
where $\alpha_{\mathrm{2D}}=1/2$ for 2D slab geometry and $\alpha_{\mathrm{3D}}=1$ for cylindrical geometry.

\begin{figure}
\centering\includegraphics[width=0.6\columnwidth]{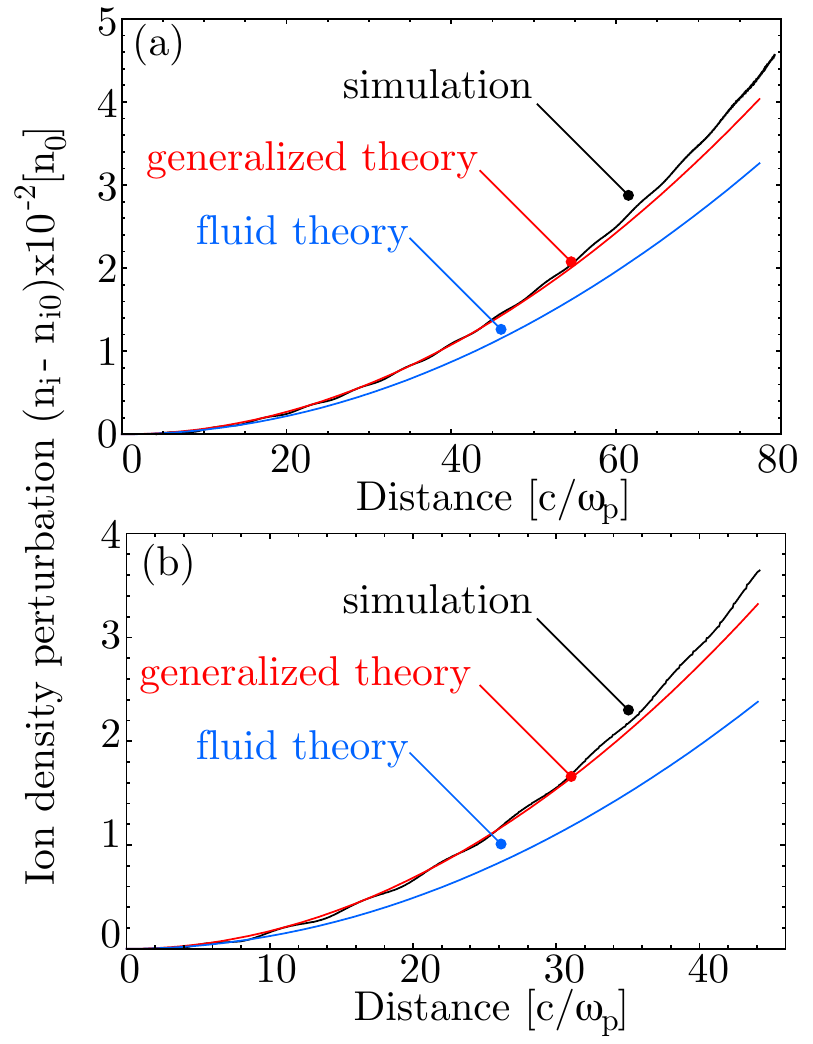}
\caption{Comparison between simulation results (black), generalized theoretical results (red) and linear theory predictions (blue) for the on-axis background plasma ion density evolution. The plasma was excited by an external electric field driver with profile given by $E_{r}^b=(m_e c\omega_p/e) E_0 x_{\perp} \exp{\left(x_{\perp}^2/\sigma_r^2\right)}$ with $\sigma_r = 0.1~c/\omega_p$ and $E_0 = 0.2$ in (a) and $E_0=0.3$ in (b).}
\label{fig:early-ionmotion}
\end{figure}

Figure~\ref{fig:early-ionmotion} shows the on-axis background plasma ion density given by simulations and theory inside a long external field driver. The numerical simulation parameters and setup are similar to those of Fig.~\ref{fig:plasma}. The external electric field profile is given by Eq.~(\ref{eq:driver}) with $\sigma_{\perp}=0.1~c/\omega_p$ and $E_0=0.2$ (Fig.~\ref{fig:early-ionmotion}a) and $E_0=0.3$ (Fig.~\ref{fig:early-ionmotion}b). Theory is given by Eq.~(\ref{eq:ni-cyl-onaxis}), where we used Eq.~(\ref{eq:amplitude-forced}) to evaluate $\hat{E}_<$. Figure~\ref{fig:early-ionmotion} also provides comparisons with linear theory results obtained by expanding $n_i$ to lowest order in $\mathcal{O}(\nabla_{r}\hat{E}_<)$ in Eqs.~(\ref{eq:amplitude-forced}) and (\ref{eq:ni-cyl-onaxis}). We find very good agreement between Eq.~(\ref{eq:ni-cyl-onaxis}) and simulation results. %
However, Fig.~\ref{fig:early-ionmotion} shows that linear theory results under-estimate $n_i$ by roughly 20\%. This is consistent with results shown in Fig.~\ref{fig:field-comp} because the average $\langle E_r \rangle$ from linear theory predictions given by Eq.~(\ref{eq:efield-forced-average-taylor}) is lower than Eq.~(\ref{eq:efield-forced-average}). For propagation distances larger than those shown in Fig.~\ref{fig:early-ionmotion} the agreement between theory and simulations becomes worse, with theory underestimating simulation results. Simulations suggest that this is due to plasma wave breaking associated with the fine scale mixing of electron trajectories, not accounted for in this model. 

To estimate the onset of the ion motion in self-modulated regimes (or in regimes resonantly driven by trains of particle bunches), we consider that the wakefield grows secularly along the beam. In this scenario, the ion motion is mainly driven by the plasma wakefields in the absence of driver (i.e. through $\langle E_>\rangle$). The wakefield amplitude ($\hat{E}_>$) at the back of the driver bunch with length $\sigma_z$, is thus:
\begin{equation}
\label{eq:smfield-average}
\hat{E}_> \equiv \hat{E}_>^{\mathrm{ave}} = \frac{\sigma_z}{2 \lambda_p} \hat{E}_>^{\mathrm{wake}}\simeq \frac{\sigma_z}{2 \lambda_p}E_r^b.
\end{equation}
Replacing Eq.~(\ref{eq:smfield-average}) in the expression for $\langle E_>\rangle$ [Eq.~(\ref{eq:efield-average})] yields the average resonantly driven transverse plasma wakefield ponderomotive force. Near the axis, the onset of the ion motion can then be estimated by determining the position ($\xi_{\mathrm{crit}}$) where on-axis ion density compression reaches an accepted value. The onset of the ion motion in the PDPWFA becomes important when $\xi_{\mathrm{crit}}/\sigma_z \simeq 1$. Assuming that the on-set for the ion motion occurs in 3D when $n_i \simeq 2 n_{i0}$ then:
\begin{equation}
\label{eq:onset-ionmotion}
\frac{\xi_{\mathrm{crit}}}{\sigma_z}= \frac{c^2}{\omega_p^2 \sigma_z^2}\left(\frac{m_i }{2 m_e Z}\right)^{1/2}\left(\frac{4 \pi m_e \omega_p^2}{e \nabla_r E_>^{\mathrm{wake}}}-\frac{3}{8\pi}\frac{e \sigma_z^2 \nabla_r E_>^{\mathrm{wake}}}{m_e c^2}\right).
\end{equation}
A typical value for the wake produced by current PDPWFA experiments is $e \nabla_r E_>^{\mathrm{wake}}/(m_e \omega_p^2)\simeq 10^{-2}$ ($n_b/n_0\simeq 10^{-2}$, where $n_b$ is the proton bunch density). Thus, in an Hydrogen plasma ($m_i/m_e=1836$ and $Z=1$) the ion motion becomes important for $\xi_{\mathrm{crit}} \gtrsim 200c/\omega_p\simeq30\lambda_p \simeq \sigma_z$. It is therefore expected that experiments currently being prepared could be affected by the ion motion if Hydrogen plasmas are used. In order to avoid the ion motion heavier ions are thus required. In this example, the role of the ion motion could be suppressed if $Z m_i/m_e\simeq 50 m_\mathrm{proton}/m_e$, where $m_{\mathrm{proton}}$ is the mass of the proton.

We can also show that the onset of the ion motion occurs sooner than plasma wave breaking due to the fine scale mixing of electron trajectories. The onset of fine scale mixing of forced plasma electron trajectories occurs after~\cite{bib:dawson_pr_1959}:
\begin{equation}
\label{eq:wb}
\xi_{\mathrm{WB}} = \frac{1}{4}\frac{\lambda_p^2}{R_{\mathrm{max}}}\frac{\mathrm{d} r_{e0}}{\mathrm{d} T(r_0)},
\end{equation}
where $R_{\mathrm{max}}$ is the amplitude of the electron oscillation, and $T(r_{e0})$ is the period of oscillation of an electron initially at $r=r_{e0}$. 
Assuming typical values for $R_{\mathrm{max}}\simeq (n_b/n_0)\sigma_r$, and using Eq.~(6) to determine $\mathrm{d}T(r_{e0})/\mathrm{d} r_{e0}\simeq (2\pi \xi/\phi^2)(\mathrm{d}\phi/\mathrm{d}r_{e0})\simeq (2\pi c/\omega_p)(e/m_e c\omega_p^2)(\mathrm{d}^2 E_r^b/\mathrm{d} r_{e0}^2) \simeq (2\pi c/\omega_p) (n_b/n_0) (1/\sigma_r)$ then $\xi_{\mathrm{WB}}\simeq (c/\omega_p) (n_0/n_b)^2$. Hence, for $n_b/n_0\simeq 10^{-2}$, $\xi_{\mathrm{WB}}\simeq 10000 c/\omega_p$, much longer than the onset of the ion motion in hydrogen.
%
In this estimate we neglected geometrical effects leading to anharmonic plasma electron oscillations. Including only geometrical effects leads to~\cite{bib:dawson_pr_1959}:
\begin{equation}
\label{eq:wbcyl}
\xi_{\mathrm{WB}} = \frac{3\pi}{\omega_p}\frac{r_{e0}^2}{R_{\mathrm{max}}^2}\left(\frac{\mathrm{d} R_{\mathrm{max}}}{\mathrm{d} r_{e0}}\right).
\end{equation}
For the typical parameters mentioned above, and taking $\mathrm{d}R_{\mathrm{max}}/\mathrm{d}r_{r0}\simeq (n_b/n_0)(1/\sigma_r)$, then $\xi_{\mathrm{WB}}\simeq 70000 c/\omega_p$ which is also much longer than the onset of the ion motion in an Hydrogen plasma. 

In order to complement our analytical findings, we performed additional 2D cylindrically symmetric simulations of the self-modulated PDPWFA. The dimensions of the simulation moving   window were $680\times8~\left(c/\omega_p\right)^2$, divided into $6800\times160$ cells with $2\times2$ electrons per cell. An initially $450~\mathrm{GeV}$ proton bunch was initialised with density profile 
given by $n_b=n_{b0}\left[1+\cos\left(\sqrt{\pi}{2}\left(z-z_0\right)/\sigma_z\right)\right]\exp\left[-r^2/(2\sigma_r^2)\right]$ for $0<z<z_0=\sigma_z \sqrt{2 \pi}$, with $\sigma_z=225.6 c/\omega_p$, $\sigma_r=0.376~c/\omega_p$ and $n_{b0}=0.0152$. For a plasma density $n_0=10^{14}~\mathrm{cm}^{-3}$ this corresponds to $\sigma_z=12~\mathrm{cm}$ and $\sigma_r=200~\mu\mathrm{m}$. 

\begin{figure}
\centering\includegraphics[width=0.6\columnwidth]{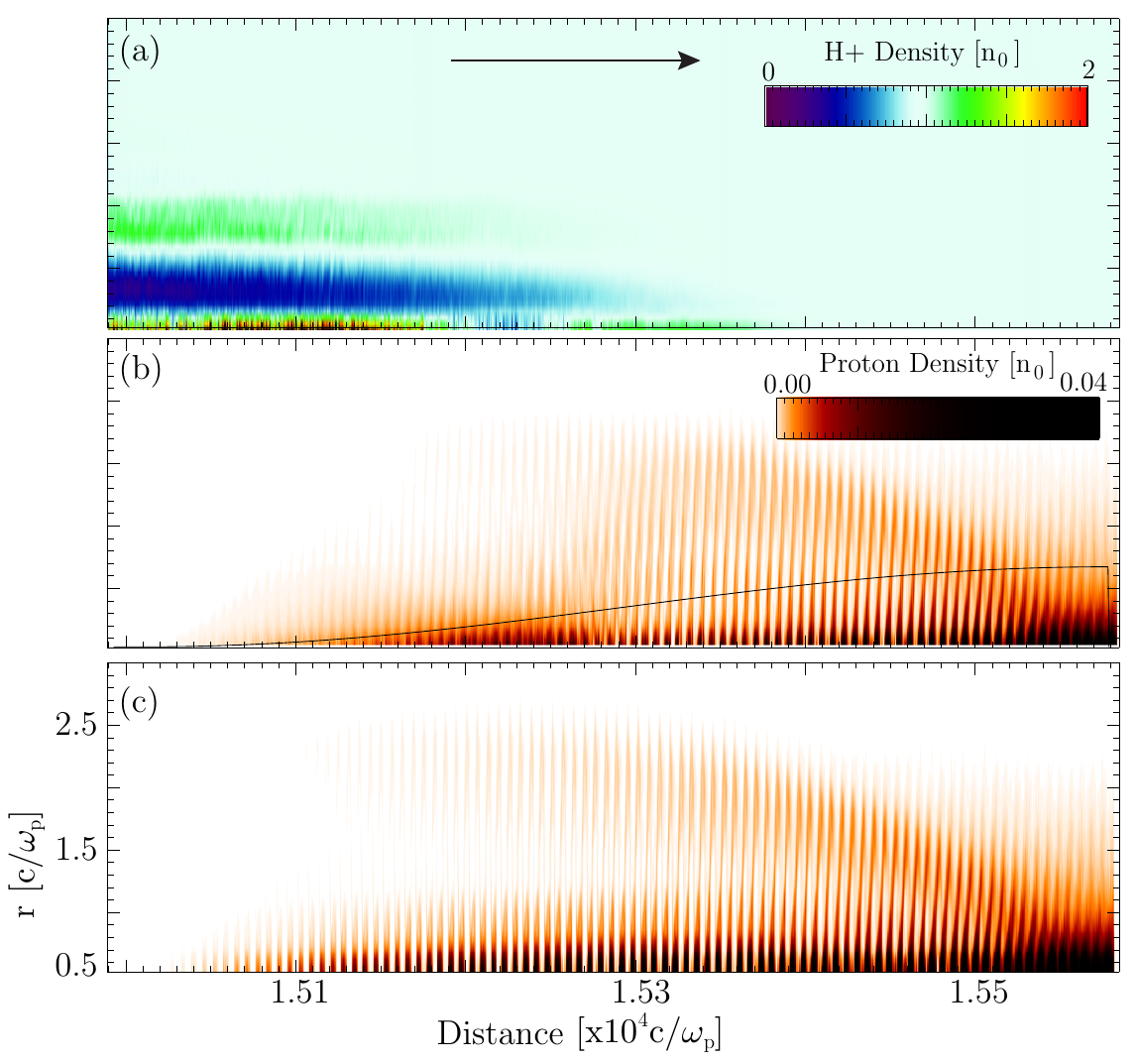}
\caption{Simulation results showing the influence of the background plasma ion dynamics on self-modulation of a long proton bunch. (a) background Hydrogen plasma ion density modulations after a propagation of almost 8 meters in the plasma. The arrow indicates the proton bunch propagation direction. (b) corresponding proton bunch self-modulated profile showing suppression of the self-modulation. (c) shows a fully self-modulated proton bunch profile in an immobile ion plasma.}
\label{fig:pdpwfa-ionmotion}
\end{figure}

Figure~\ref{fig:pdpwfa-ionmotion} shows the impact of the background ion motion in the self-modulation. Fig.~\ref{fig:pdpwfa-ionmotion}a illustrates the dynamics of background Hydrogen plasma ions after almost 8 meters of propagation. It shows that background plasma ions accumulate on-axis where the ion density significantly exceeds $n_{i0}$. The ion motion has a significant impact on the trajectories of plasma electrons, strongly reducing the wave breaking time [c.f. Eqs.~(\ref{eq:wb}) and (\ref{eq:wbcyl})]. The turbulent electron flow that appears once wave breaking occurs neutralizes space charge fields. As a result, the radial focusing force that drives the self-modulation instability vanishes~\cite{bib:vieira_prl_2012} leading to self-modulation suppression. Figure~\ref{fig:pdpwfa-ionmotion}b shows these effects where self-modulation is suppressed for the last half of the bunch. Self-modulation suppression is also evident from comparing bunch self-modulation in Hydrogen (Fig.~\ref{fig:pdpwfa-ionmotion}b) and in an immobile ion plasma (\ref{fig:pdpwfa-ionmotion}c).

\begin{figure}
\centering\includegraphics[width=0.8\columnwidth]{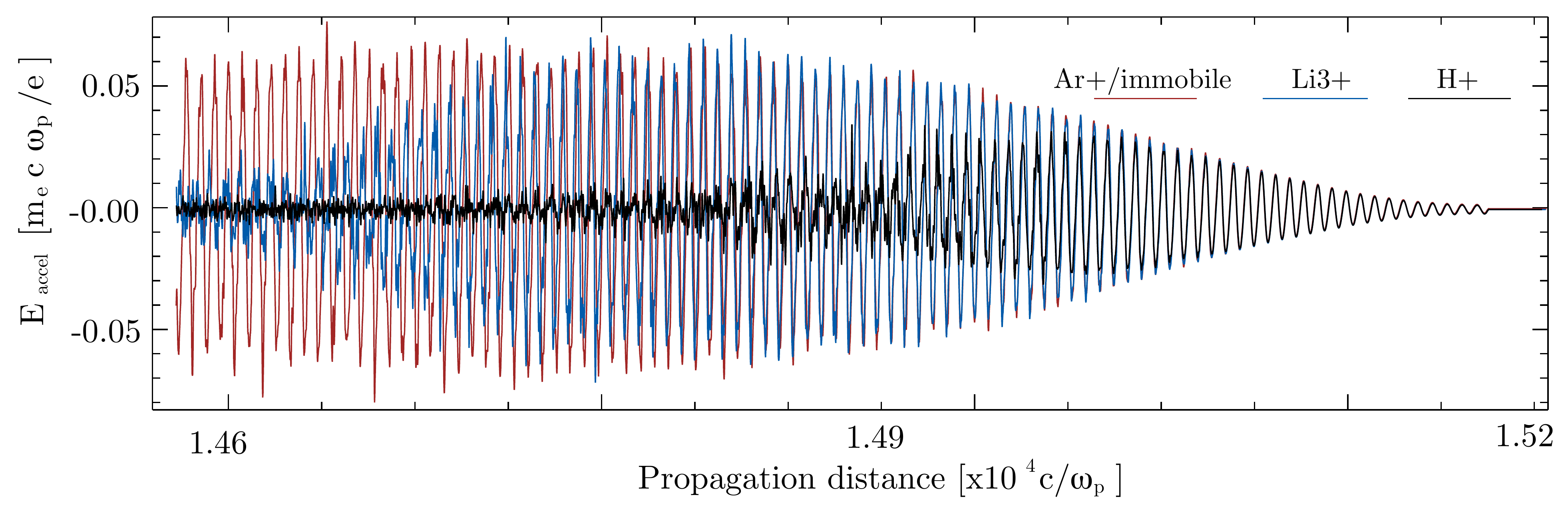}
\caption{Lineouts of the on-axis accelerating gradients after almost 8 m of propagation in the PDPWFA using Hydrogen (black line), Lithium (blue line) and Argon or Immobile ions (red line).}
\label{fig:pdpwfa-eaccel}
\end{figure}

In addition to the suppression of plasma focusing forces, the accelerating fields can also drop significantly when the ions move. This is shown in Fig.~\ref{fig:pdpwfa-eaccel} which compares the accelerating fields for plasmas with different ion masses  after the proton bunch propagated 8 meters in the plasma. The accelerating fields drop significantly near the head of the bunch in Hydrogen. In Lithium the accelerating gradients decay later along the bunch. In this case, the maximum fields are similar to those of Immobile or Argon ions for which the accelerating fields are maintained throughout the whole bunch. However, the maximum charge that could be accelerated may be lower than when using Argon because the total energy stored in the accelerating fields is lower.

\section{Conclusions}
\label{sec:sec5}
In this work we derived a kinetic model for the excitation of transverse wakefields based on J. Dawson's plasma sheet model~\cite{bib:dawson_pr_1959}. Our work generalises linear theory results in the limit of narrow wakefields. We derived expressions for the ponderomotive force of the plasma wave acting on plasma ions. PIC simulations confirmed theoretical predictions demonstrating the importance of ion motion in the conditions of the PDPWFA. Mitigating the ion motion requires using heavier ions. In addition, this work showed numerical simulations where fine scale mixing of plasma electron trajectories and wavebreaking occurs even in the limit where the amplitude of the electron trajectories is small.

\acknowledgements
The authors acknowledge fruitful discussions with Prof. A. Pukhov. This work was supported by the European Research Council (ERC-2010-AdG Grant No. 267841) and by FCT (Portugal) through Grants Nos. EXPL/FIS-PLA/0834/2012 and PTDC/FIS/111720/2009. Other support was given through U. S. DOE Grants Nos. DE-FC02-07ER41500 and DE-FG02- 92ER40727 and NSF Grants Nos. NSF PHY-0904039 and PHY-0936266. Simulations were done on the IST Cluster at IST, Jaguar supercomputer under INCITE. We acknowledge PRACE from awarding access to resource SuperMUC based in Germany at Leibniz research centre.


\begin{thebibliography}{40}
\bibitem{bib:chen_prl_1985} P. Chen \emph{et al.} Phys. Rev. Lettl. \textbf{54}, 693 (1985). 
\bibitem{bib:tajima_prl_1979} T. Tajima and J. M. Dawson, Phys. Rev. Lett. \textbf{43}, 267 (1979).
\bibitem{bib:beatwave} M. N. Rosenbluth and C. S. Liu, Phys. Rev. Lett. \textbf{29}, 701-705 (1972); B. I. Cohen, A. Kaufman, K. Watson, Phys. Rev. Lett. \textbf{29} 581 (1972); C.E. Clayton, K. A. Marsh, A. Dyson, M. Everett, A. Lal, W.P. Leemans, R. Williams, C. Joshi, Phys. Rev. Lett. \textbf{70} 37 (1992); C. Joshi, W.B. Mori, T. Katsouleas, J. M. Dawson, J. M. Kindel, D. W. Forslund, Nature \textbf{311} 525 (1984); P. Gibbon, Phys. Fluids B \textbf{2}, 2196-2208 (1990); N. E. Andreev, L. M. Gorbunov, V. I. Kirsanov, A. A. Pogosova, and R. R. Ramazashvili, JETP Lett. \textbf{55}, 571-576 (1992).
\bibitem{bib:lsmi} E. Esarey, J. Krall, P. Sprangle Phys. Rev. Lett. \textbf{72} 2887 (1994); K. Nakajima \emph{et al} Phys. Rev. Lett. \textbf{74} 4428 (1995); P. Sprangle, E. Esarey, J. Krall, and A. Ting, Phys. Rev. Lett. \textbf{69}, 2200-2203 (1992); T. M. Antonsen, Jr. and P. Mora, Phys. Rev. Lett. \textbf{69}, 2204-2207 (1992)
\bibitem{bib:etrain} E. Esarey, P. Sprangle, J. Krall, A. Ting, IEEE-TPS \textbf{24}, 252 (1996) and references therein; P. Muggli, V. Yakimenko, M. Babzien, E. Kallos, and K. P. Kusche, Phys. Rev. Lett. \textbf{101}, 054801 (2008); 
\bibitem{bib:ltrain} D. Umstadter, E. Esarey, J. Kim, Phys. Rev. Lett. \textbf{72} 1224 (1994); S. Dalla, M. Lontano, Phys. Rev. E \textbf{49} R1819 (1994); G. Bonnaud, D. Teychenne, and J.-L. Bobin, Phys. Rev. E \textbf{50}, R36-R39 (1994); S. Dalla and M. Lontano, Plasma Phys. Control. Fusion \textbf{36},1987-2002 (1994); D. Umstadter, J. Kim, E. Esarey, E. Dodd, and T. Neubert, Phys. Rev. E \textbf{51}, 3484-3497 (1995); S. Kalmykov and G. Shvets, Phys. Plasmas \textbf{13}, 056707 (2006); S. Kalmykov and G. Shvets, Phys. Plasmas AIP Conf. Proc. \textbf{877}, 395-401 (2006). 
\bibitem{bib:lu_2006} J. Rosensweig \emph{et al.}, Phys. Rev. A \textbf{44}, R6189-6192 (1991); W. Lu \emph{et al.}, Phys. Rev. Lett. \textbf{96}, 165005 (2006); W. Lu \emph{et al.}, Phys. Rev. ST Accel. Beams \textbf{10}, 061301 (2007); A. Pukhov, J. Meyer-ter-Vehn App. Physics B \textbf{74} 355 (2002).
\bibitem{bib:lwfa} W. P. Leemans, B. Nagler, A. J. Gonsalves, Cs. Toth, K. Nakamura, C. G. R. Geddes, E. Esarey, C. B. Schoreder, S. M. Hooker, Nat. Physics \textbf{2} 696 (2006); X. Wang, R. Zgadzaj, N. Fazel, Z. Li, S. A. Yi, X. Zhang, W. Henderson,	 Y.-Y. Chang, R. Korzekwa, H.-E. Tsai, C.-H. Pai, H. Quevedo, G. Dyer, E. Gaul, M. Martinez, A. C. Bernstein, T. Borger, M. Spinks, M. Donovan, V. Khudik, G. Shvets, T. Ditmire,  M. C. Downer, Nature Communications ~\textbf{4} 1988 (2013); S. Kneip et al., Phys. Rev. Lett. 103, 035002 (2009); C. E. Clayton et al., Phys. Rev. Lett. 105, 105003 (2010); S. Banerjee, N. D. Powers, V. Ramanathan, I. Ghebregziabher, K. J. Brown, C. M. Maharjan, S. Chen, A. Beck, E. Lefebvre, S. Y. Kalmykov, B. A. Shadwick, and D. P. Umstadter, Phys. Plasmas \textbf{19}, 056703 (2012).
\bibitem{bib:pwfa}I. Blumenfeld \emph{et al.}, Nature \textbf{445}, 741 (2007).
\bibitem{bib:caldwell_natphys_2009} A. Caldwell \emph{et al.}, Nat. Phys. 5, 363 (2009).
\bibitem{bib:kumar_prl_2010} N. Kumar, A. Pukhov, and K. Lotov, Phys. Rev. Lett. \textbf{104} 255003 (2010).
\bibitem{bib:mori_ieee_1997} W.B. Mori, IEEE J. Quantum Electron. \textbf{33}, 1942 (1997).
\bibitem{bib:silva_aip_2006} L. O. Silva, AIP Conf. Proc. 856, 109 (2006).
\bibitem{bib:schroeder_prl_2011} C. B. Schroeder \emph{et al.}, Phys. Rev. Lett. \textbf{107} 145002 (2011).
\bibitem{bib:pukhov_prl_2011} A. Pukhov \emph{et al.}, Phys. Rev. Lett. \textbf{107} 145003 (2011).
\bibitem{bib:vieira_pop_2012} J. Vieira, Y. Fang, W.B. Mori, L.O. Silva, P. Muggli, Phys. Plasmas \textbf{19} 063105 (2012).
\bibitem{bib:fiberlaser} C. Jauregui, J. Limper, A. T\"unermann Nat. Photonics 20 October (2013); M. Fermann, I. Hartl Nat. Photonics 20 October (2013).
\bibitem{ionmotion} E. V. Chizhonkov and L. M. Gorbunov, Russ. J. Numer. Anal. Math. Modelling, \textbf{16}(3), 235-246 (2001); L. M. Gorbunov, P. Mora, and A. A. Solodov, Phys. Rev. Lett. \textbf{86}(15), 3332-3335 (2001); L. M. Gorbunov, P. Mora, and A. A. Solodov, Phys. Plasmas~\textbf{10}, 1124-1134 (2003). 
\bibitem{bib:vieira_prl_2012} J. Vieira, R.A. Fonseca, W.B. Mori, L.O. Silva, Phys. Rev. Lett. \textbf{109} 145005 (2012).
\bibitem{bib:rosensweig_prl_2005} J. B. Rosenzweig,A.M. Cook, A. Scott, M.C. Thompson, R. Yoder, Phys. Rev. Lett. \textbf{95} 195002 (2005).
\bibitem{bib:muggli_ion} R. Gholizadeh, T. Katsouleas, P. Muggli, C. Huang, and W. Mori, Phys. Rev. Lett. \textbf{104} 155001 (2010).
\bibitem{bib:dawson_pr_1959} J.M. Dawson Phys. Rev. \textbf{113}, 383 (1959).
\bibitem{bib:osiris} R. A. Fonseca \emph{et al.}, Lect. Notes Comp. Sci. vol. 2331/2002, (Springer Berlin / Heidelberg, (2002); R.A. Fonseca \emph{et al.}, Plasma Phys. Control. Fusion \textbf{50}, 124034 (2008).
\bibitem{bib:lu_pop_2005} W. Lu, C. Huang, M.M. Zhou, W.B. Mori, T.Katsouleas, Phys. Plasmas \textbf{12} 063101 (2005).
\end{thebibliography}
\end{document}